\documentclass[aps,prl,twocolumn,showpacs,nofootinbib,superscriptaddress]{revtex4-1}
\pdfoutput=1

\usepackage{graphicx}
\usepackage{bm}
\usepackage{amssymb,amsmath,latexsym}
\usepackage{color}
\definecolor{darkblue}{RGB}{0,0,196}
\definecolor{darkred}{RGB}{196,0,0}
\usepackage[colorlinks=true,linktocpage=true,linkcolor=darkblue,citecolor=darkblue,urlcolor=darkblue]{hyperref}

\DeclareMathOperator{\arctanh}{arctanh}

\def\be{\begin{equation}}
\def\ee{\end{equation}}
\def\ba{\begin{eqnarray}}
\def\ea{\end{eqnarray}}

\begin{document}

\title{3+1d quasiparticle anisotropic hydrodynamics for ultrarelativistic heavy-ion collisions}

\author{Mubarak Alqahtani} 
\affiliation{Department of Physics, Kent State University, Kent, OH 44242 United States}

\author{Mohammad Nopoush} 
\affiliation{Department of Physics, Kent State University, Kent, OH 44242 United States}

\author{Radoslaw Ryblewski}
\affiliation{The H. Niewodnicza\'nski Institute of Nuclear Physics, Polish Academy of Sciences, PL-31342 Krak\'ow, Poland}

\author{Michael Strickland} 
\affiliation{Department of Physics, Kent State University, Kent, OH 44242 United States}

\begin{abstract}
We present the first comparisons of experimental data with phenomenological results from 3+1d quasiparticle anisotropic hydrodynamics (aHydroQP).  We compare charged-hadron multiplicity, identified-particle spectra, identified-particle average transverse momentum, charged-particle elliptic flow, and identified-particle elliptic flow produced in LHC 2.76 TeV Pb+Pb collisions.  The dynamical equations used for the hydrodynamic stage utilize non-conformal aHydroQP.  The resulting aHydroQP framework naturally includes both shear and bulk viscous effects in addition to higher-order non-linear transport coefficients.  The 3+1d aHydroQP evolution obtained is self-consistently converted to hadrons using anisotropic Cooper-Frye freezeout performed on a fixed-energy-density hypersurface.  The final production and decays of the primordial hadrons are modeled using a customized version of THERMINATOR 2.  In this first study, we utilized smooth Glauber-type initial conditions and a single effective freeze-out temperature $T_{\rm FO} = 130$ MeV with all hadronic species in full chemical equilibrium.  With this rather simple setup, we find a very good description of many heavy-ion observables.
\end{abstract}

\date{\today}

\pacs{12.38.Mh, 24.10.Nz, 25.75.Ld, 47.75.+f}

\keywords{Quark-gluon plasma, Relativistic heavy-ion collisions, Anisotropic hydrodynamics, Equation of state, Boltzmann equation}

\maketitle

%%%%%%%%%%%%%%%%%%%%%%%%%%%%%%%%%%%%%%%%%%%%%%%%%%%%%%%%%%%

Ultrarelativistic heavy-ion collision experiments at the Relativistic Heavy Ion Collider and Large Hadron Collider (LHC) were designed to create and study the quark-gluon plasma (QGP).   Relativistic hydrodynamics has been quite successful in describing the collective behavior observed in high-energy heavy-ion collisions \cite{Huovinen:2001cy,Hirano:2002ds,Kolb:2003dz} and the current focus of the relativistic hydrodynamics community is on further improvements of the models to include e.g.~bulk viscous effects and higher-order transport coefficients \cite{Muronga:2001zk, Muronga:2003ta, Muronga:2004sf, Heinz:2005bw, Baier:2006um, Romatschke:2007mq, Song:2007fn, Dusling:2007gi, Song:2007ux, Baier:2007ix, Luzum:2008cw, Song:2008hj, Heinz:2009xj, Denicol:2010tr, Denicol:2010xn, Schenke:2010rr, Schenke:2011tv, Bozek:2011wa, Niemi:2011ix, Niemi:2012ry, Bozek:2012qs, Denicol:2012cn,Denicol:2014vaa,Ryu:2015vwa,Tinti:2016bav} (see \cite{Heinz:2013th, Gale:2013da, Jeon:2016uym} for recent reviews).  The goal of the relativistic viscous hydrodynamics program is to constrain key properties of the QGP such as its initial energy density, initial pressure anisotropies, shear viscosity, bulk viscosity, etc.~and to also provide the soft-background evolution necessary to compute QGP-modification of hard probes such as jets and heavy quark bound states.

One of the issues faced by practitioners of traditional second-order viscous hydrodynamics approaches is that, at early times after the nuclear impact, the QGP possesses a high degree of momentum-space anisotropy in the fluid local rest frame, ${\cal P}_T/{\cal P}_L \gg 1$.  The magnitude of the resulting momentum-space anisotropy is large at early times after the initial nuclear impact and also near the transverse/longitudinal ``edges'' of the QGP at all times.  In these spacetime regions, traditional viscous hydrodynamics is being pushed to its limits, resulting in potentially negative total pressures and violations of positivity of the one-particle distribution function~\cite{Strickland:2014pga}.  

As a way to address these problems, it was suggested that one should reorganize the expansion of the one-particle distribution function around a leading-order form which possesses intrinsic momentum-space anisotropies but still guarantees positivity~\cite{Florkowski:2010cf,Martinez:2010sc}.  This method has become known as anisotropic hydrodynamics (aHydro).  Since the two original papers~\cite{Florkowski:2010cf,Martinez:2010sc}, there has been a great deal of progress in aHydro \cite{Martinez:2012tu,Ryblewski:2012rr,Bazow:2013ifa,Tinti:2013vba,Nopoush:2014pfa,Tinti:2015xwa,Bazow:2015cha,Strickland:2015utc,Alqahtani:2015qja,Molnar:2016vvu,Molnar:2016gwq,Alqahtani:2016rth} including applications to cold atomic gases near the unitary limit \cite{Bluhm:2015raa,Bluhm:2015bzi}.  In parallel, there have been efforts to construct exact solutions to the Boltzmann equation in some simple cases which can be used to test the efficacy of various dissipative hydrodynamics approaches, and it has been shown that aHydro most accurately reproduces all known exact solutions, even in the limit of very large $\eta/s$ and/or initial momentum-space anisotropy \cite{Florkowski:2013lza,Florkowski:2013lya,Denicol:2014tha,Denicol:2014xca,Nopoush:2014qba,Heinz:2015gka,Molnar:2016gwq}.  

A recent focus of research has been on turning aHydro into a practical phenomenological tool with a realistic equation of state (EoS) and self-consistent anisotropic hadronic freeze-out.  In this paper, we present the first comparisons of experimental data with phenomenological results obtained using (1) generalized 3+1d aHydro including three momentum-space anisotropy parameters in the underlying distribution function, (2) the quasiparticle aHydro (aHydroQP) method for implementing a realistic EoS \cite{Alqahtani:2015qja, Alqahtani:2016rth, Alqahtani:2016ayv} and (3) anisotropic Cooper-Frye freezeout~\cite{Nopoush:2015yga,Alqahtani:2016rth} using the same distribution form as was assumed for the dynamical equations.  All previous phenomenological applications of aHydro have relied on the approximate conformal factorization of the energy-momentum tensor, see e.g.~\cite{Martinez:2012tu,Ryblewski:2012rr,Nopoush:2016qas,Strickland:2016ezq}, and/or have used isotropic freezeout \cite{Ryblewski:2012rr}.  For modeling the primordial hadron production and subsequent hadronic decays we use a customized version of THERMINATOR 2 which has been modified to accept ellipsoidally anisotropic distribution functions~\cite{Chojnacki:2011hb}.  

\textit{ 1.~Model:} 
In aHydro, the leading-order one-particle distribution function is assumed to be of generalized Romatschke-Strickland form \cite{Romatschke:2003ms,Martinez:2012tu,Nopoush:2014pfa}
\be
f(x,p) = f_{\rm iso}\!\left(\frac{1}{\lambda}\sqrt{p_\mu \Xi^{\mu\nu} p_\nu}\right) ,
\label{eq:genf}
\ee
where $\lambda$ is an energy scale which resembles the temperature in the anisotropic distribution, $\Xi^{\mu\nu} \equiv u^\mu u^\nu + \xi^{\mu\nu} - \Phi\Delta^{\mu\nu}$ is the anisotropy tensor, $\xi^{\mu\nu}$ obeys $u_\mu \xi^{\mu\nu}=0$ and ${\xi^\mu}_\mu = 0$, $\Phi$ is the bulk degree of freedom, and $\Delta^{\mu\nu} = g^{\mu\nu} - u^\mu u^\nu$ is the transverse projector. Since $\xi^{\mu\nu}$ is traceless and orthogonal to $u^\mu$, there are five independent components.  In this work, we assume that $\xi^{\mu\nu}$ is diagonal, $\xi^{\mu\nu} = {\rm diag}(0,{\boldsymbol \xi})$, in which case it has only two independent degrees of freedom.  Taken together with $\Phi$, this gives three independent degrees of freedom which map to three ellipsoidal anisotropies in momentum space.  It is assumed that the function $f_{\rm iso}(x)$ is a thermal distribution function which can be identified as a Fermi-Dirac, Bose-Einstein, or classical Boltzmann distribution. Herein, we use the Boltzmann distribution when computing the various moment integrals entering the aHydroQP equations of motion and we assume that the fluid chemical potentials are zero at all times.  When freezing-out to specific hadron types, however, we use the quantum distribution appropriate for each particle type.

%%%%%%%%%%%%%%%%%%%%%%%%%%%%%%%%%%%%%%%%%%%%%%%%%%%%%%%%%%%
\begin{figure}[t!]
\vspace{3.5mm}
\includegraphics[width=.95\linewidth]{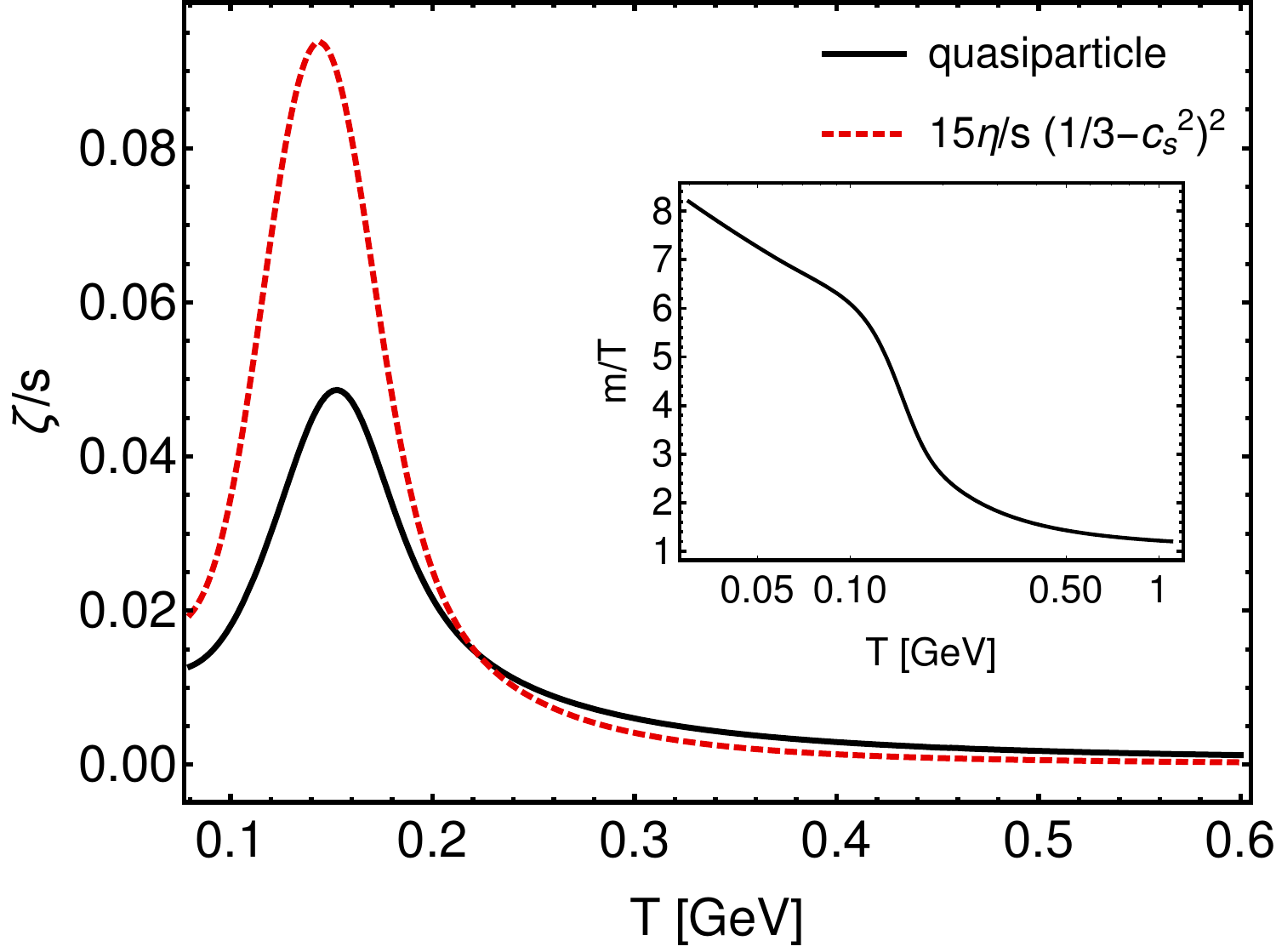}
\caption{The scaled bulk viscosity obtained using a quasiparticle model with a single temperature-dependent mass (black solid line) \cite{Romatschke:2011qp,Tinti:2016bav} and for comparison $\zeta/s = 15 \eta/s \, (1/3 - c_s^2)^2$, which is a frequently used small-mass limit expression (red dashed line).  The inset shows $m/T$ extracted by fitting to lattice data \cite{Borsanyi:2010cj} for the QCD entropy density.}
\label{fig:zeta}
\end{figure}
%%%%%%%%%%%%%%%%%%%%%%%%%%%%%%%%%%%%%%%%%%%%%%%%%%%%%%%%%%%

In order to obtain the dynamical equations necessary, we consider a system of quasiparticles with energy-density-dependent masses.  In this case, the Boltzmann equation is~\cite{Jeon:1995zm}
\be
p^\mu\partial_\mu f+\frac{1}{2}\partial_i m^2\partial^i_{(p)}f =-{\cal C}[f]\,,
\label{eq:boltzeq}
\ee
with `$i$' labeling spatial indices and  ${\cal C}[f]$ being the collisional kernel which, herein, we treat in relaxation-time approximation. 
In order to conserve energy-momentum and maintain thermodynamic consistency in equilibrium, one must introduce an additional degree of freedom to the energy-momentum tensor, $T^{\mu\nu} = T^{\mu\nu}_{\rm kinetic}+B g^{\mu\nu}$, where, in general, $B$ is a function of all system parameters and $g^{\mu\nu}$ is the metric tensor~\cite{gorenstein1995gluon,Jeon:1995zm,Romatschke:2011qp,Alqahtani:2015qja,Alqahtani:2016rth}.  By taking momentum-moments of the quasiparticle Boltzmann equation (\ref{eq:boltzeq}), one can obtain a system of partial differential equations for the ellipsoidal anisotropy parameters $\boldsymbol\xi$, the scale parameter $\lambda$, and the fluid four-velocity $u^\mu$.  These dynamical equations form the basis of aHydroQP \cite{Alqahtani:2015qja}.  The three anisotropies encode the effects of both shear and bulk viscous corrections and, as usual, the fluid four-velocity is normalized to unity.  

%%%%%%%%%%%%%%%%%%%%%%%%%%%%%%%%%%%%%%%%%%%%%%%%%%%%%%%%%%%
\begin{figure}[t!]
\includegraphics[width=0.925\linewidth]{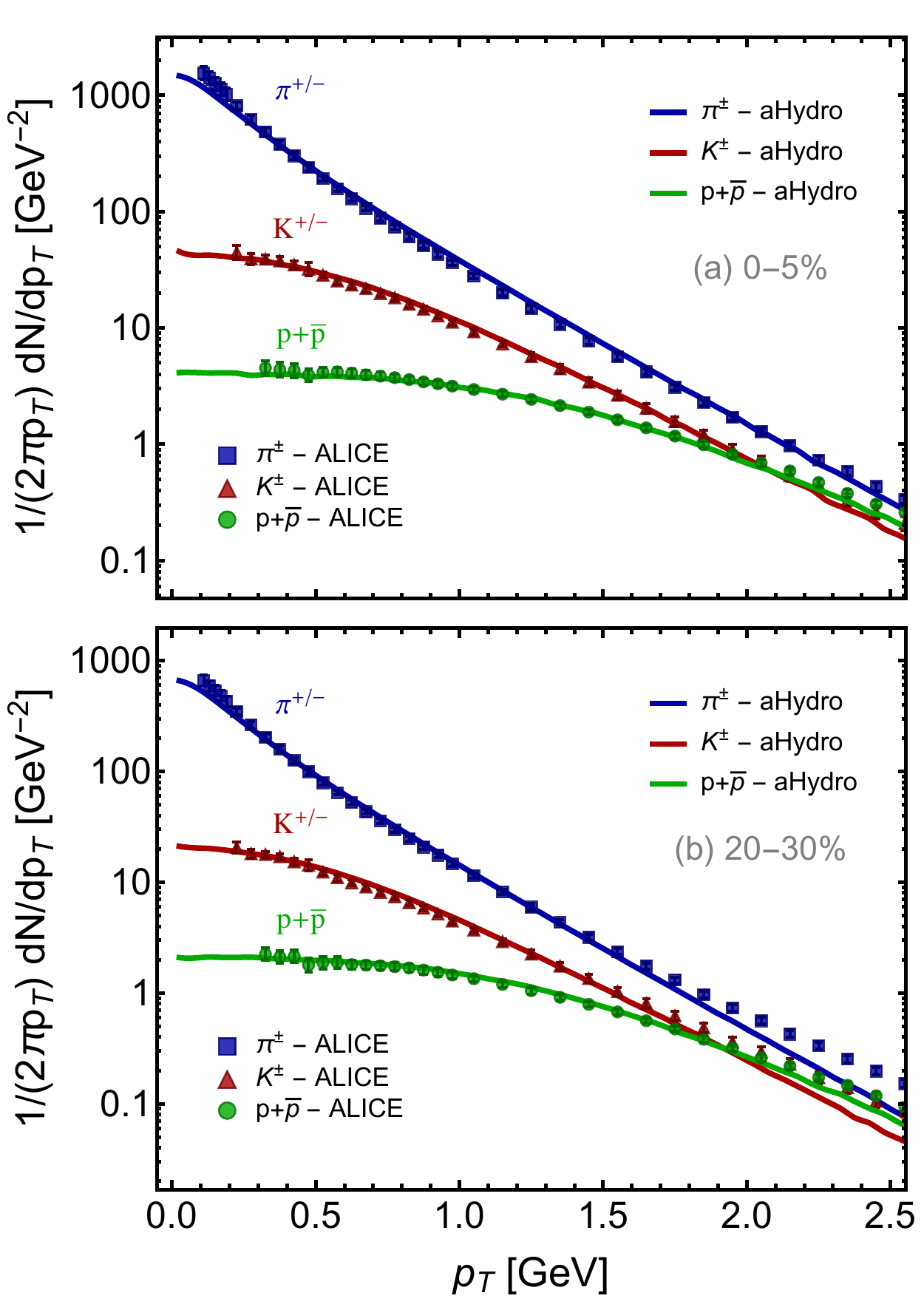}
\caption{Spectra of $\pi^\pm$, $K^\pm$, and $p+\bar{p}$ as a function of $p_T$ for centrality classes  0-5$\%$ and 20-30$\%$.  All results are for 2.76 TeV Pb+Pb collisions.  Data shown are from the ALICE collaboration \cite{Abelev:2013vea}.}
\label{fig:spectra}
\end{figure}
%%%%%%%%%%%%%%%%%%%%%%%%%%%%%%%%%%%%%%%%%%%%%%%%%%%%%%%%%%%

%%%%%%%%%%%%%%%%%%%%%%%%%%%%%%%%%%%%%%%%%%%%%%%%%%%%%%%%%%%
\begin{figure*}[t!]
\centerline{
\hspace{-1.5mm}
\includegraphics[width=.33\linewidth]{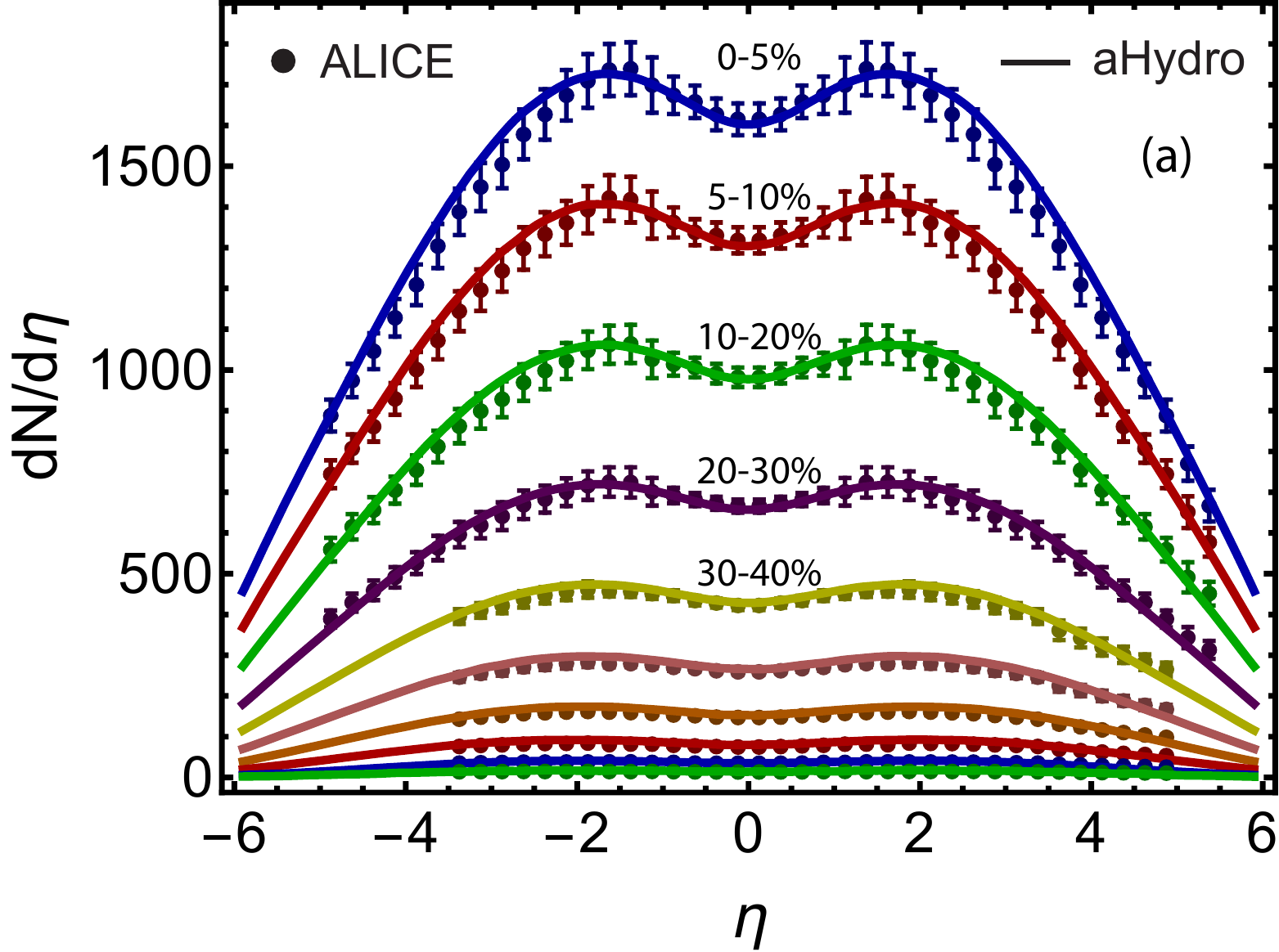}
\hspace{-0.5mm}
\includegraphics[width=.32\linewidth]{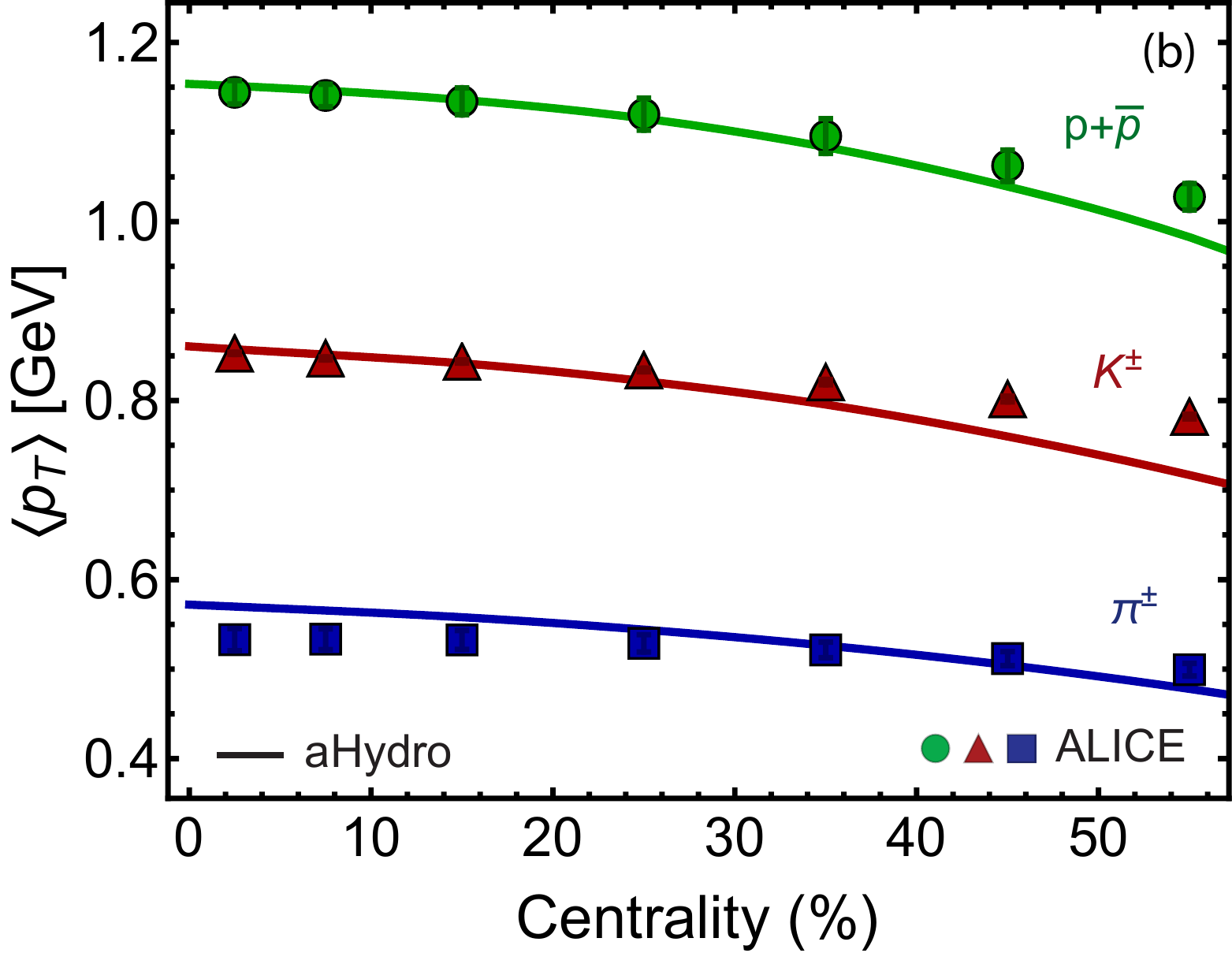}
%\hspace{1mm}
\includegraphics[width=.33\linewidth]{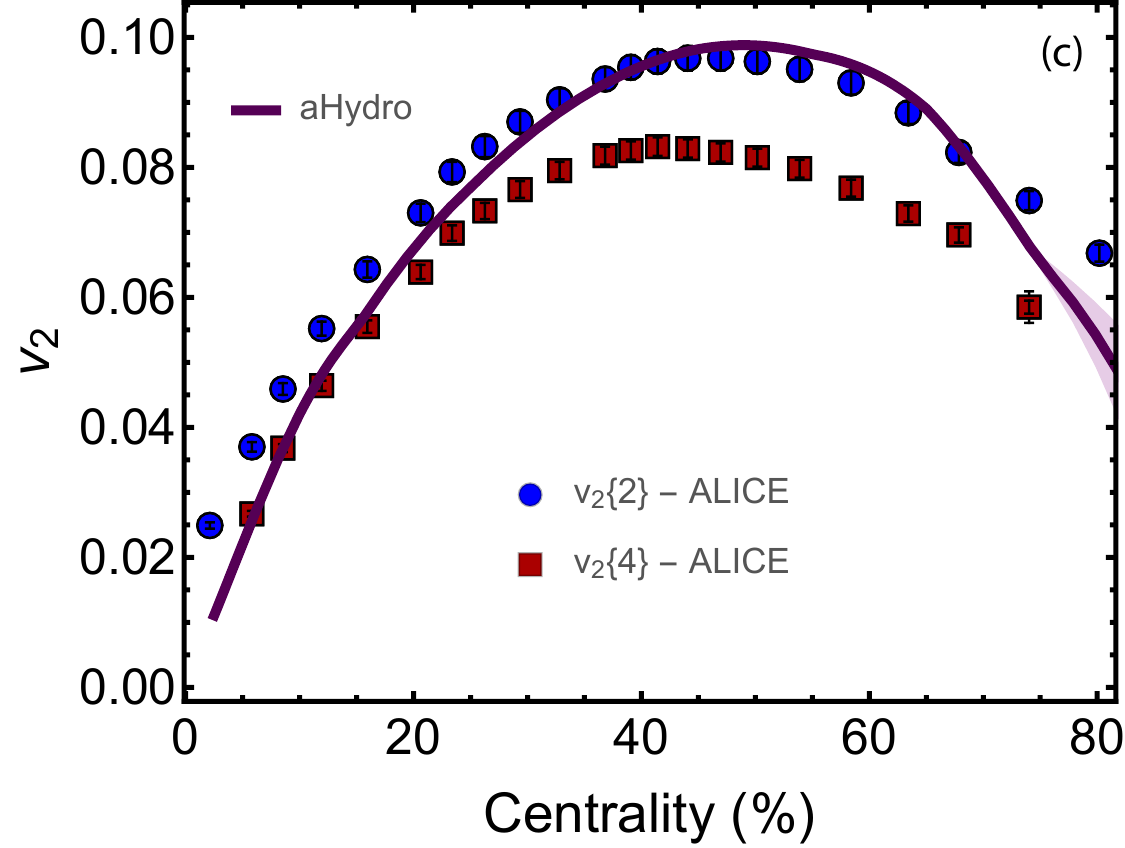}
}
\caption{Three panels showing:  (a) the charged-hadron multiplicity in different centrality classes as a function of pseudorapidity; (b) the average transverse momentum of pions, kaons, and protons as a function of centrality; and (c) the integrated $v_2$ for charged hadrons as a function of centrality ($0.2 < p_T < 3$ GeV, $\eta < 0.8$).  All results are for 2.76 TeV Pb+Pb collisions.  Data in panels (a)-(c) are from the ALICE collaboration Refs.~\cite{Abbas:2013bpa,Adam:2015kda}, \cite{Abelev:2013vea}, and \cite{Abelev:2014mda}, respectively. }
\label{fig:combined}
\end{figure*}
%%%%%%%%%%%%%%%%%%%%%%%%%%%%%%%%%%%%%%%%%%%%%%%%%%%%%%%%%%%

Using the quasiparticle setup, one can extract the bulk viscosity in the near-equilibrium limit.  In Refs.~\cite{Romatschke:2011qp} and \cite{Tinti:2016bav} one can find expressions for the bulk viscosity to entropy density $\zeta/s$ in Eqs.~(4.4) and (45), respectively.  When evaluated, both expressions give the same result for a system of quasiparticles with a temperature-dependent mass.  The result is plotted as a black solid line in Fig.~\ref{fig:zeta}.  In addition to this exact analytic expression, which is valid for all values of $m/T$, we plot an often-used small-mass expansion result, \mbox{$\zeta/s = 15 \eta/s \, (1/3 - c_s^2)^2$}, for purposes of comparison as a red dashed line.  For both curves we assumed that $\eta/s = 0.159$.

In the inset of Fig.~\ref{fig:zeta} we plot the extracted value of $m/T$ obtained by fitting to the Wuppertal-Budapest continuum-extrapolated results for the QCD entropy density \cite{Alqahtani:2015qja}.  As can be seen from the inset, at small temperatures, the value of $m/T$ necessary to fit the lattice data \cite{Borsanyi:2010cj} is not small, invalidating commonly used small-mass approximations.  As Fig.~\ref{fig:zeta} demonstrates, the quasiparticle model used herein has a finite bulk viscosity to entropy density ratio which peaks in the vicinity of the phase transition from QGP to a hadronic gas; however, the magnitude of the peak is much smaller than many other phenomenologically used ans\"atze for $\zeta/s$.  For example, in Ref.~\cite{Ryu:2015vwa} the authors have a peak value of $\zeta/s$ which is approximately $0.3$.

\textit{ 2.~Results  and Discussions:}  
In this paper we present comparisons of our aHydroQP results with \mbox{$\sqrt{s_{NN}}$ = 2.76 TeV} Pb+Pb collision data available from the ALICE collaboration.  For our initial condition we take the system to be isotropic in momentum space with zero transverse flow and Bjorken flow in the longitudinal direction.  In the transverse plane, the initial energy density is computed from a linear combination of smooth Glauber wounded-nucleon and binary-collision profiles with a binary mixing factor of $\alpha = 0.15$.  In the longitudinal direction, we used a ``tilted'' profile with a central plateau and Gaussian ``wings'' resulting in a profile function of the form $\rho(\varsigma) \equiv \exp \left[ - (\varsigma - \Delta \varsigma)^2/(2 \sigma_\varsigma^2) \, \Theta (|\varsigma| - \Delta \varsigma) \right]$, with \mbox{$\varsigma = \arctanh(z/t)$} being spatial rapidity.  The parameters entering the longitudinal profile function were fitted to the pseudorapidity distribution of charged hadrons with the results being $\Delta\varsigma = 2.3$ and $\sigma_{\varsigma} = 1.6$.  The first quantity sets the width of the central plateau and the second sets the width of the Gaussian ``wings''.  

The resulting initial energy density at a given transverse position ${\bf x}_\perp$ and spatial rapidity $\varsigma$ was computed using ${\cal E} \propto (1-\alpha) \rho(\varsigma) \left[ W_A({\bf x}_\perp) g(\varsigma) + W_B({\bf x}_\perp) g(-\varsigma)\right] + \alpha \rho(\varsigma) C({\bf x}_\perp)$, where $W_{A,B}({\bf x}_\perp)$ is the wounded nucleon density for nuclei $A$ and $B$, $C({\bf x}_\perp)$ is the binary collision density, and $g(\varsigma)$ is the ``tilt function''.  The tilt function $g(\varsigma) = 0$ if $\varsigma < -y_N$, $g(\varsigma) = (\varsigma+y_N)/(2y_N)$ if $-y_N \leq \varsigma \leq y_N$, and $g(\varsigma)=1$ if $\varsigma > y_N$ where $y_N = \log(2\sqrt{s_{NN}}/(m_p + m_n))$ is the nucleon momentum rapidity \cite{Bozek:2010bi}.

For all results presented herein, we solved the aHydroQP dynamical equations on a $64^3$ lattice with lattice spacings $\Delta x = \Delta y = 0.5$ fm and \mbox{$\Delta \varsigma$ = 0.375}.  We computed spatial derivatives using fourth-order centered-differences and, for temporal updates, we used fourth-order Runge-Kutta with step size of $\Delta\tau = 0.02$ fm/c and a weighted-LAX smoother to regulate potential numerical instabilities associated with the centered-differences scheme~\cite{Martinez:2012tu}.  We started the aHydroQP evolution at $\tau_0 = 0.25$ fm/c and ended it when the highest effective temperature in the three-volume was sufficiently below the freeze-out temperature.

After running the full 3+1d evolution of the system using aHydroQP, we extracted a fixed energy-density freeze-out hypersurface corresponding to a given effective temperature.  The fluid anisotropy tensor and scale parameter were assumed to be the same for all hadronic species and we additionally assumed that all produced hadrons were in chemical equilibrium.  The aHydroQP distribution function parameters on the freezeout hypersurface were fed into a customized version of THERMINATOR 2 which uses Monte-Carlo sampling to generate final hadronic configurations.  Once the primordial hadrons were sampled in this manner, the subsequent hadronic decays proceeded as usual.  In the plots shown herein, we used between 7,400 and 36,200 hadronic events, depending on the centrality class and the target observable, e.g.~for centrality classes in which we show identified-particle $v_2(p_T)$, more hadronic events were used in order to increase statistics.  In all plots, the statistical uncertainty of our model results associated with hadronic Monte-Carlo sampling is indicated by a shaded band surrounding the central line, which indicates the hadronic event-averaged value.

To fix the remaining model parameters, we used scans in the initial central temperature $T_0$, the freezeout temperature $T_{\rm FO}$, and $\eta/s$, where the latter was assumed to be a temperature-independent constant.  The theoretical predictions resulting from this scan were compared to experimental data from the ALICE collaboration for the differential spectra of pions, kaons, and protons in both the 0-5\% and 30-40\% centrality classes.  The fitting error was minimized across species, with equal weighting for the three particle types.  The parameters obtained from this procedure were $T_0 = 600$ MeV, $\eta/s = 0.159$, and \mbox{$T_{\rm FO} = 130$ MeV}.  Herein, $T_0$ is the initial temperature which would be obtained in a perfectly central collision at ${\bf x}=0$.  The resulting fit to the spectra that emerged in the 0-5\% and 20-30\% centrality classes is shown in Fig.~\ref{fig:spectra}.  As can be seen from Fig.~\ref{fig:spectra}, the resulting spectra fits are quite good, allowing for a simultaneous description of the pion, kaon, and proton spectra.  Note that, for pions, the model slightly underpredicts the pion spectrum at low transverse momentum.  This discrepancy is similar to what is observed in other hydrodynamic models.

Using the parameters determined by the procedure outlined above, we then proceeded to calculate other observables.  In Fig.~\ref{fig:combined}, we present three panels which show (a) the charged-hadron multiplicity in different centrality classes as a function of pseudorapidity; (b) the average transverse momentum of pions, kaons, and protons as a function of centrality; and (c) the integrated $v_2$ for charged hadrons as a function of centrality.  In each panel, we compare to data reported by the ALICE collaboration.  As can be seen from panel (a), our model is able to describe the charged hadron multiplicity as a function pseudorapidity quite well in all centrality classes.  From panel (b) we see that the model is also able to reproduce the average $p_T$ of pions, kaons, and protons quite well.  The fit quality achieved is similar to Ref.~\cite{Ryu:2015vwa}, however, we note that in our model the peak value of $\zeta/s$ is substantially smaller than what was assumed in Ref.~\cite{Ryu:2015vwa}.  Turning to panel (c) we compare our model predictions computed using the geometrical $v_2 \sim \langle \cos(2\phi) \rangle$ for all charged hadrons with ALICE results obtained using second- and fourth-order cumulants $v_2\{2\}$ and $v_2\{4\}$.  As we can see from panel (c),  the model underestimates $v_2$ for very central collisions.  This is to be expected since we did not include event-by-event fluctuations in the initial condition.  That being said, we see that the qualitative behavior of $v_2$ as a function of centrality is well reproduced.

\begin{figure}[t!]
\includegraphics[width=0.95\linewidth]{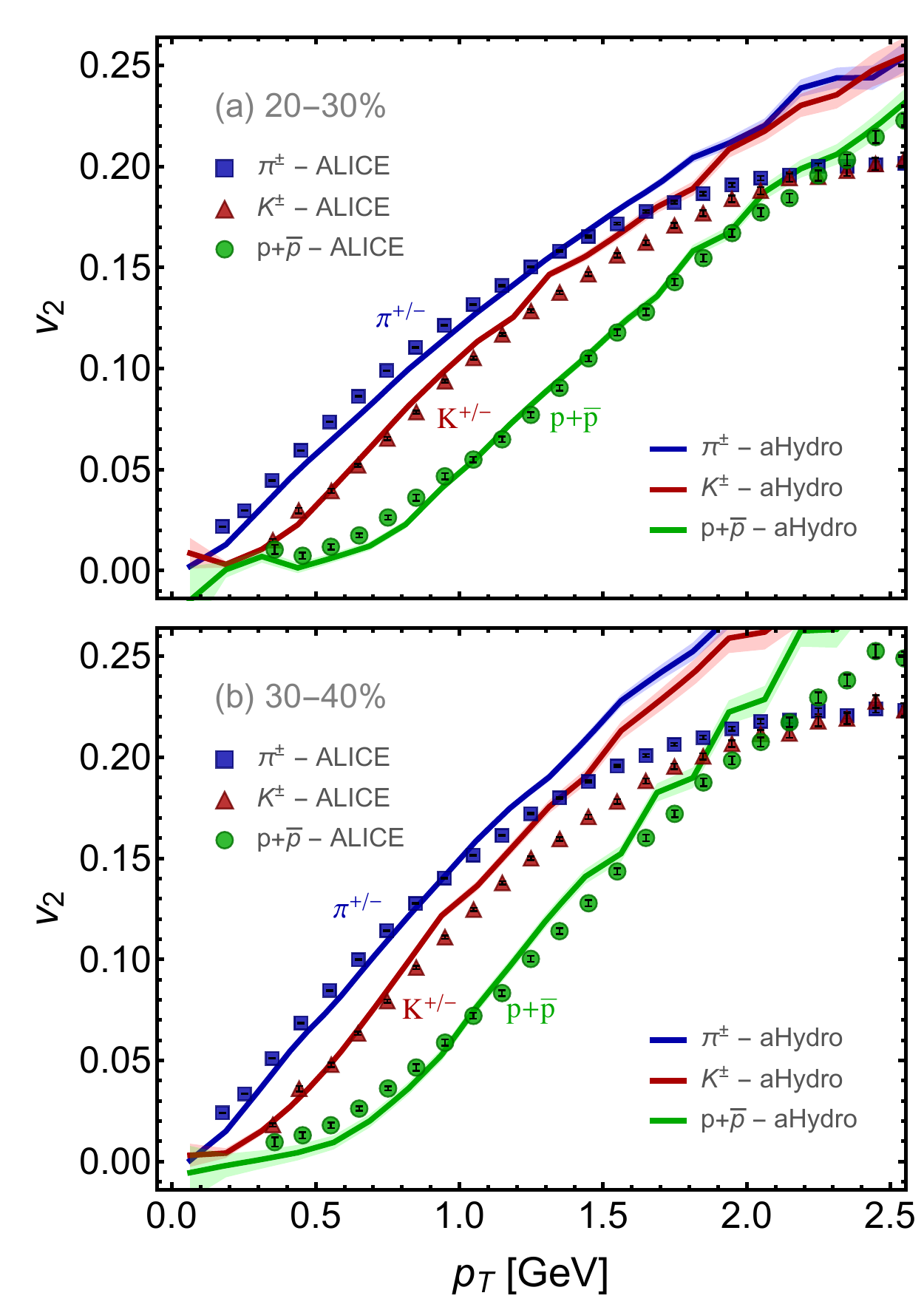}
\caption{The elliptic flow coefficient for identified hadrons as a function of  $p_T$ for centrality classes  20-30$\%$ and  30-40$\%$. All results and data are for 2.76 TeV Pb+Pb collisions.  Data shown are from the ALICE collaboration and were extracted using the scalar product method~\cite{Abelev:2014pua}.}
\label{fig:v2}
\end{figure}

Finally, turning to Fig.~\ref{fig:v2} we present two panels which compare our model predictions for the identified-particle $v_2(p_T)$ with experimental data from the ALICE collaboration.  The top and bottom panels show the results obtained in the 20-30\% and 30-40\% centrality classes, respectively.  As can be seen from these panels, the model provides a very good description of the identified-particle elliptic flow.  In the 20-30\% centrality class, the model is in good agreement with the pion, kaon, and proton data out to $p_T \sim$ 1.5, 1.5, and 2.5 GeV, respectively.  In the 30-40\% centrality class, the model is in good agreement with the pion, kaon, and proton data out to $p_T \sim$ 1, 1, and 2 GeV, respectively.  In order to improve the agreement between theory and data, it would seem that one has to, at the very least, relax the assumption of a temperature independent $\eta/s$.  Since high-momentum hadrons are produced significantly at early times after the collision when the energy density is high, it is natural to expect that elliptic flow would be reduced at high $p_T$ since the effective shear viscosity would be larger. 

%%%%%%%%%%%%%%%%%%%%%%%%%%%%%%%%%%%%%%%%%%%%%%%%%%%%%%%%%%%
%%%%%%%%%%%%%%%%%%%%%%%%%%%%%%%%%%%%%%%%%%%%%%%%%%%%%%%%%%%

\textit{ 3.~Conclusions and Outlook:}
In this paper, we have provided the first phenomenological comparisons of aHydroQP with LHC experimental data, which includes (1)~generalized aHydroQP including three momentum-space anisotropy parameters in the underlying distribution function, (2)~the quasiparticle method for implementing a realistic EoS and (3)~anisotropic Cooper-Frye freezeout using the same distribution form as was assumed for the dynamical equations.  For modeling the primordial hadron production and subsequent hadronic decays we used a customized version of THERMINATOR 2 which has been modified to accept ellipsoidally anisotropic distribution functions.  For this initial application, we assumed smooth Glauber initial conditions which were a linear combination of wounded-nucleon and binary-collision profiles.  We further assumed that the system was initially isotropic in momentum space.  With these assumptions, we performed a parameter scan and, through comparisons of the identified-particle spectra emerging from the model and experimental data, we were able to find a best fit with $T_0 = 600$ MeV, $\eta/s = 0.159$, and \mbox{$T_{\rm FO} = 130$ MeV}.  With this small set of parameters we were able to obtain good agreement between the model and experimental data for the identified-particle spectra, the identified-particle average transverse momentum as a function of centrality, the charged-hadron multiplicity as a function of pseudorapidity, the charged-particle $v_2$ as a function of centrality, and the identified-particle $v_2$ as a function of transverse momentum.  We note, in particular, that we were able to obtain a good description of the average transverse momentum of pions, kaons, and protons with a much smaller peak value for the bulk viscosity to entropy density ratio than previous studies (see e.g.~Ref.~\cite{Ryu:2015vwa}).  This suggests that there is a fair amount of hydrodynamical model variation in statements about the magnitude of the bulk viscosity in the QGP.

This study provides the first solid evidence that it is possible to apply aHydroQP to obtain a successful phenomenological description of the QGP.  Of course, this study is only the first major step.  Looking forward, it is necessary to include realistic fluctuating initial conditions, temperature-dependent shear viscosity to entropy density ratio, realistic initial momentum-space anisotropy profiles (tied to the fluctuating initial conditions), more realistic collisional kernels, etc.  These are all currently in progress and, based on the results obtained herein using a somewhat simple setup, we are quite optimistic that aHydroQP can be used as a reliable modeling tool in the future.  

Looking beyond the phenomenological applications, we emphasize that the aHydroQP formalism used herein represents an important step forward in the self-consistent implementation of both large shear corrections and non-conformal effects in relativistic dissipative hydrodynamics.  Recent works have shown that, in the context of second-order viscous hydrodynamics, the self-consistent incorporation of the temperature-dependence of the quasiparticle mass (vHydroQP) results in important modifications to QGP transport coefficients, particularly at low temperatures~\cite{Tinti:2016bav}.  Additionally, Ref.~\cite{Tinti:2016bav} demonstrated that prior aHydroQP results published in Ref.~\cite{Alqahtani:2015qja} agree very well with the vHydroQP evolution and that both quasiparticle formalisms result in a qualitatively different evolution of the bulk viscous correction compared to existing approaches which ignore the temperature dependence of the quasiparticle mass.  The phenomenological results obtained herein suggest that, if non-conformal aspects are more carefully taken into account, one can obtain a very good description of many key heavy-ion observables.

%%%%%%%%%%%%%%%%%%%%%%%%%%%%%%%%%%%%%%%%%%%%%%%%%%%%%%%%%%%
\acknowledgements{{\it Acknowledgements:} We thank D. Keane for useful comments.  M.~Alqahtani was supported by a PhD fellowship from the University of Dammam, Saudi Arabia.  M.~Nopush and M.~Strickland were supported by the U.S. Department of Energy, Office of Science, Office of Nuclear Physics under Award No. DE-SC0013470.  R.~Ryblewski was supported by the Polish National Science Center grant No. DEC-2012/07/D/ST2/02125.}
%%%%%%%%%%%%%%%%%%%%%%%%%%%%%%%%%%%%%%%%%%%%%%%%%%%%%%%%%%%

\bibliography{3p1ahydro}

\end{document}